\documentclass[lettersize,journal]{IEEEtran}
\usepackage{amsmath,amsfonts}
\usepackage{algorithmic}
\usepackage{algorithm}
\usepackage{array}
\usepackage[caption=false,font=normalsize,labelfont=sf,textfont=sf]{subfig}
\usepackage{textcomp}
\usepackage{stfloats}
\usepackage{url}
\usepackage{verbatim}
\usepackage{graphicx}
\usepackage{cite}
\usepackage{color}
\usepackage{amsthm}

\newtheorem{remark}{Remark}

\newtheorem{corollary}{Corollary}
\begin{document}

\title{Performance Analysis of XL-MIMO with Rotary and Movable Antennas for High-speed Railway}

\author{Wenhui Yi,~Jiayi Zhang,~\IEEEmembership{Senior Member,~IEEE},~Zhe Wang, Huahua Xiao,~Bo Ai,~\IEEEmembership{Fellow,~IEEE}
\vspace{-1cm}
\thanks{W. Yi, J. Zhang, Z. Wang, and B. Ai are with the School of Electronics and Information Engineering, Beijing Jiaotong University, Beijing 100044, P. R. China. H. Xiao is ZTE Corporation, Shenzhen, P. R. China.}}

\maketitle

\begin{abstract}
The rotary and movable antennas (ROMA) technology is efficient in enhancing wireless network capacity by adjusting both the antenna spacing and three-dimensional (3D) rotation of antenna surfaces, based on the spatial distribution of users and channel statistics. Applying ROMA to high-speed rail (HSR) wireless communications can significantly improve system performance in terms of array gain and spatial multiplexing. However, the rapidly changing channel conditions in HSR scenarios present challenges for ROMA configuration. In this correspondence, we propose a analytical framework for configuring ROMA-based extremely large-scale multiple- input-multiple-
output (XL-MIMO) system in HSR scenarios based on spatial correlation. First, we develop a localization model based on a mobility-aware near-field beam training algorithm to determine the real-time position of the train relay antennas. Next, we derive the expression for channel orthogonality and antenna spacing based on the spatial correlation matrix, and obtain the optimal antenna spacing when the transceiver panels are aligned in parallel. Moreover, we propose an optimization algorithm for the rotation angle of the transceiver panels, leveraging the differential evolution method, to determine the optimal angle. Finally, numerical results are provided to validate the computational results and optimization algorithm. 
\end{abstract}

\vspace{-0.5cm}
\section{Introduction}
Deploying extremely large-scale multiple-
input-multiple-output (XL-MIMO)  systems in high-speed railway (HSR) environments significantly improves degrees of freedom (DoFs) and spectral efficiency (SE), which can substantially enhance wireless network coverage\cite{ref1,ref2}. However, this multi-antenna technique essentially represents MIMO with fixed-position antennas. While increasing the number of fixed-position antennas improves performance, it also leads to higher hardware costs and greater power consumption. Furthermore, in a wireless network utilizing fixed-position antennas (FPA), the allocation of antenna resources cannot be dynamically adjusted based on the spatial distribution of user channels, beyond the capabilities of traditional adaptive MIMO processing\cite{ref7}. As a result, fixed-position MIMO systems face inherent limitations. To fully exploit the spatial variations in wireless channels at base stations (BS) and wireless terminals, the rotary and movable antennas (ROMA) technique has been proposed as a novel and cost-effective solution for enhancing wireless network performance.

ROMA is an emerging next-generation multiple antenna technology, which can flexibly adjust the antenna spacing and array rotation angles of the transceiver. The fundamental idea is to enhance the spatial freedom and channel capacity of the MIMO system without increasing the number of antennas, achieved by adjusting the transceiver antenna units and the 3D geometric features of the entire array. There are several recent studies that adopt similar ideas. In \cite{MA}, the authors analyzed the performance of a system in which the Access Point (AP) is able to both move and rotate. In\cite{intro2}, the authors optimized the average network capacity for a random number of users located at random positions by jointly adjusting the 3D positions and rotations of multiple 6DMA surfaces. In \cite{intro3}, the authors proposed a fluid antenna system in which the physical position of an antenna can be switched freely to one of the $N$ positions over a fixed-length line space to pick up the strongest signal in the manner of traditional selection diversity. Compared to previous studies, the ROMA technique enables adjusting the geometric characteristics of transceiver antenna units and the overall array, addressing the growing communication demands of HSR scenarios. While continuously rotating the ROMA surface offers maximum flexibility and the greatest capacity enhancement, the rapid movement of the HSR leads to frequent changes in channel conditions, making it challenging to determine the optimal rotation angle of the ROMA surface at any given moment.

In this correspondence, we deploy an XL-MIMO system with ROMA in HSR scenario. We then analyze and optimize the system based on spatial correlation, considering factors such as spatial freedom, system capacity, and other relevant metrics. First, we develop a localization model using mobility-aware near-field beam training to predict the XL-MIMO system's position over time. Next, we calculate the antenna spacing that ensures spatial orthogonality of the channel, based on the channel's spatial correlation matrix. Finally, we determine the optimal rotation angles for both the transceiver and the end panels in 3D space using a differential optimization algorithm. Simulation results validate the effectiveness of the optimal antenna spacing expression and the rotation angle optimization algorithm.

 \vspace{-0.4cm}
\section{System Model}
We consider a downlink XL-MIMO system for HSR scenario, wherein the transmitter and receiver are UPAs with $M=M_H\times M_V$ and $N=N_H\times N_V$ antenna  
elements, respectively, where $M_H(N_H)$ is the number of the antenna elements in the horizontal direction and $M_V(N_V)$ is the number of elements in the vertical direction. The uniform antenna spacing across the vertical and horizontal direction between adjacent Tx antennas as $d_{tv}$ and $d_{th}$, respectively, and between Rx antennas as $d_{rv}$, and $d_{rh}$. As shown in Fig. 1, the geometric relationship between two opposing UPAs is defined by four 3D rotation angles: $\alpha_1$, $\beta_1$, $\alpha_2$, and $\beta_2$, representing the transmitter's rotation about the $x$-axis, tilt relative to the $z$-axis, and the receiver's rotation about the $x$-axis and tilt relative to the $z$-axis, respectively. The transmitter is centered in $\mathbf{r}^t=[0,0,0]^T$, while the receiver is centered in $\mathbf{r}^r=[x_0,y_0,z_0]^T$.  Hence, the distance between the center points of the antenna arrays is 
\vspace{-0.2cm}
\begin{equation}
\label{1}
    D=|\mathbf{r}^t-\mathbf{r}^r|=\sqrt{{x_0}^2+{y_0}^2+{z_0}^2}.
    \vspace{-0.2cm}
\end{equation}
\begin{figure}[!t]
\centering
\vspace{-0.4cm}
\includegraphics[scale=0.5]{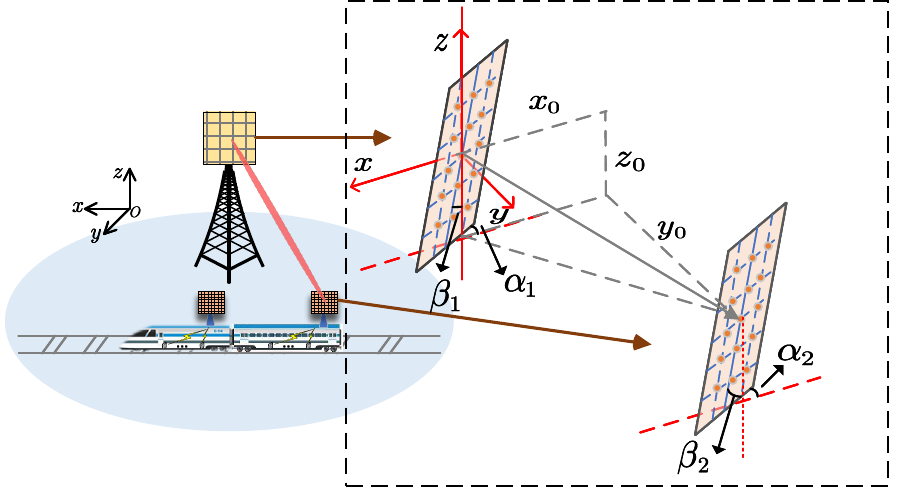}
\vspace{-0.4cm}
\label{fig_first_case}
\caption{ROMA-enabled HSR XL-MIMO communication systems. }
\label{fig_sim}
\vspace{-0.6cm}
\end{figure}
The positions of the $m$-th transmit antenna element and the $n$-th receive antenna element are denoted by $\mathbf{r}_{m}^{t}=[ \varDelta _{xm}^{t},\varDelta _{ym}^{t},\varDelta _{zm}^{t} ] ^T$ and $\mathbf{r}_{n}^{r}=[ x_0+\varDelta _{xn}^{r},y_0+\varDelta _{yn}^{r},z_0+\varDelta _{zn}^{r} ] ^T$, respectively. $\varDelta _{an}^{r}$ and $\varDelta _{am}^{t}$ for the $a$-axis with $a=\{x,y,z\}$ are the coordinates respect to the center of the transmitter and receiver, respectively. $\varDelta _{an}^{r}$ and $\varDelta _{am}^{t}$ can be shown as 
\vspace{-0.2cm}
\begin{equation}
\footnotesize
 \begin{cases}
 \label{TX}
	\varDelta _{xm}^{t}=( m_1-\frac{M_H-1}{2}) d_{th}\cos \alpha _1-( m_2-\frac{M_V-1}{2}) d_{tv}\sin \beta _1\sin \alpha _1\\
	\varDelta _{ym}^{t}=( m_1-\frac{M_H-1}{2} ) d_{th}\sin \alpha _1+( m_2-\frac{M_V-1}{2}) d_{tv}\sin \beta _1\cos \alpha _1\\
	\varDelta _{zm}^{t}=( m_2-\frac{M_V-1}{2}) d_{tv}\cos \beta _1\\
\end{cases}
\end{equation}
\begin{equation}
\footnotesize
\begin{cases}
\label{RX}
	\varDelta _{xn}^{r}=( n_1-\frac{N_H-1}{2} ) d_{rh}\cos \alpha _2-( n_2-\frac{N_V-1}{2} ) d_{rv}\sin \beta _2\sin \alpha _2\\
	\varDelta _{yn}^{r}=( n_1-\frac{N_H-1}{2} ) d_{rh}\sin \alpha _2+( n_2-\frac{N_V-1}{2} ) d_{rv}\sin \beta _2\cos \alpha _2\\
	\varDelta _{zn}^{r}=( n_2-\frac{N_V-1}{2} ) d_{rv}\cos \beta _2\\
\end{cases}
\end{equation}
where $m_1=\mathrm{mod}( m-1,M_H) $ and $m_2=\lfloor ( m-1 ) /M_H \rfloor $ are the row and column indexes of the $m$-th transmit antenna, and $n_1=\mathrm{mod}( n-1,N_H) $ and $n_2=\lfloor ( n-1 ) /N_H \rfloor $ are the row and column indexes of the $n$-th receive antenna\cite{ref4}.

The channel matrix between the transmitter and the receiver is $\mathbf{H}\in \mathbb{C} ^{N\times M}$. Since we consider HSR communication in the open scenario, the power of LoS path is much higher than those of the non-LoS (NLoS) paths. Accordingly, the channel matrix between the $m$-th transmitting antenna and the $n$-th receiving antenna can be modeled as 
\begin{equation}
\label{4}
    \left[ \mathbf{H} \right] _{nm}=\frac{e^{-jk\left| \mathbf{r}_{m}^{t}-\mathbf{r}_{n}^{r} \right|}}{4\pi \left| \mathbf{r}_{m}^{t}-\mathbf{r}_{n}^{r} \right|},
    \vspace{-0.2cm}
\end{equation}
where $k=c/f$ is the wavenumber with $c$ is the speed of the light, $f=f_c+f_d$ is the propagation frequency, $f_c$ is the carrier frequency, and $f_d=\frac{f_c}{c}\frac{( \mathbf{r}_{n}^{r}-\mathbf{r}_{m}^{t} ) \mathbf{v}}{| \mathbf{r}_{m}^{t}-\mathbf{r}_{n}^{r} |}$ is the Doppler frequency offset, and $\mathbf{v}=[v,0,0]^T$ is the velocity of the train.  Hence, the channel correlation matrix is given by $ \mathbf{G}=\mathbf{H}^*\mathbf{H}\in \mathbb{C} ^{M\times M}$. According to the considered system model, the channel capacity is given as follows:
\vspace{-0.2cm}
\begin{equation}
C=\sum_{i=1}^R{\log _2\left( 1+\frac{P_i}{\sigma ^2}\lambda _i \right)},
\end{equation}
where $\lambda _i$ is the $i$-th largest eigenvalue of $\mathbf{G}$, $R$ is the rank of $\mathbf{G}$, $\sigma ^2$ is the additive white Gaussian noise power, and $P_i$ is the power allocated to the $i$-th communication mode. In the high SNR regime and LoS conditions, as considered in this system, the capacity is maximized when $\mathbf{H}$ is an orthography matrix, rank of $\mathbf{G}$ is $R=M$ and the $R$ eigenvalues have the same magnitude. Therefore, we need to fulfill the following condition to ensure the matrix $\mathbf{H}$ is orthogonal\cite{MA1}:
\vspace{-0.3cm}
\begin{equation}
\label{6}
\small 
\mathbf{G}\left( u,v \right) =\sum_{n=1}^N{\left[ \mathbf{H}\left( n,u \right) \right] ^*\mathbf{H}\left( n,v \right)}=0\,\, \forall u\ne v=1,2,...,M.
\vspace{-0.2cm}
\end{equation}

\vspace{-0.1cm}
\section{ROMA-HSR XL-MIMO Systems}
In this section, we describe our proposed framework for optimizing the deployment of antennas for UPA-based XL-MIMO HSR communication systems. Firstly, the moving receiver is localized with the predictive beam training algorithm mentioned in \cite{beamtraining}; then we analyze the conditions that make the channel matrix $\mathbf{H}$ orthogonal; finally, the channel capacity is maximized by optimizing the antenna configuration.

\vspace{-0.4cm}
\subsection{Localization Model}\label{AA}
We construct the localization model of the user's location based on the predictive beam training model of our previous work \cite{beamtraining}. Considering the speed stability and the small curvature of the track during operation, the channel parameters can be obtained from the first two near-field beam training after the receiver enters the sender's coverage area. And these parameters can be used to predict the train's subsequent motion state while enabling real-time error correction. We define the receiver position  at time $t\in \left\{ 0,\varDelta t,...T \right\} $ as $\left( \theta _t, R_t \right) $, so we can obtain the speed and position of the receiver at that moment as
\vspace{-0.1cm}
\setcounter{equation}{6}
\begin{equation}
\footnotesize
\label{7}
\begin{cases}
	\hat{v}=\frac{R_0\cos \theta _0-R_{\varDelta t}\cos \theta _{\varDelta t}}{\varDelta t}\\
	( \hat{\theta} _t,\hat{R}_t ) =( \mathrm{a}\tan ( \frac{R_0\sin \theta _0}{R_0\cos \theta _0-vt} ) ,\sqrt{( vt ) ^2-2vtR_0\cos \theta _0+R_{0}^{2}} )\\
\end{cases},
\end{equation}
where $(\hat{\theta}_t, \hat{R}_t)=(\mathrm{arc}\tan \frac{y_0}{x_0},\sqrt{x_{0}^{2}+y_{0}^{2}})$ is the receiver's position at the moment $t$ in the form of angle and distance, $\varDelta t$ is the time interval between the first and second beam training, and $\left[ 0, T \right] $ is the time range when the train passes through the coverage area of the BS. 

Based on the principle that near-field beam training can be accurate to both angle and distance dimensions, this localization scheme improves the existing near-field hierarchical beam training and achieves position prediction for mobile users targeting HSR scenarios with low-frequency beam training, and at the same time can correct the error to a certain extent. This scheme provides the basis of user location for subsequent channel correlation analysis.

\vspace{-0.5cm}
\subsection{Channel Correlation Model}

Based on the system model shown in Fig. 1, the apertures of the antenna arrays are sufficiently small compared with the distance between their center points. Therefore, the amplitude of \eqref{4} changes slowly and it can be approximated as $1/D$. Moreover, the phase in \eqref{4} is very sensitive to the variations of $| \mathbf{r}_{m}^{t}-\mathbf{r}_{n}^{r}|$. By denoting $f(x)=2x_0\varDelta _{xn}^{r}-2x_0\varDelta _{xm}^{t}+( \varDelta _{xn}^{r}) ^2-2\varDelta _{xn}^{r}\varDelta _{xm}^{t}
+( \varDelta _{xm}^{t}) ^2$ and $g(x)=2x_0\varDelta _{xn}^{r}-2x_0\varDelta _{xm}^{t}+( \varDelta _{xn}^{r}) ^2-2\varDelta _{xn}^{r}\varDelta _{xm}^{t}$, the distance $| \mathbf{r}_{m}^{t}-\mathbf{r}_{n}^{r}|$ in the phase term can be approximated as 
\vspace{-0.2cm}
\begin{equation}
\label{8}
\footnotesize
\begin{aligned}
d_{nm}&=|\mathbf{r}_{m}^{t}-\mathbf{r}_{n}^{r}| \\
&=\sqrt{D^2+\rho \left( \mathbf{r}_{m}^{t}-\mathbf{r}_{n}^{r} \right)} \\
&\approx D\left[ 1+\frac{\rho \left( \mathbf{r}_{m}^{t}-\mathbf{r}_{n}^{r} \right)}{2D^2}-\frac{\rho ^2\left( \mathbf{r}_{m}^{t}-\mathbf{r}_{n}^{r} \right)}{8D^4} \right],
\end{aligned}
\end{equation}
where 
\begin{equation}
\label{9}
\footnotesize
\begin{aligned}
&\rho \left( \mathbf{r}_{m}^{t}-\mathbf{r}_{n}^{r} \right) =f(x)+f(y)+f(z),
\end{aligned}
\end{equation}
\begin{equation}
\label{10}
\footnotesize
\begin{aligned}
&\rho ^2\left( \mathbf{r}_{m}^{t}-\mathbf{r}_{n}^{r} \right) \approx (g(x)+g(y)+g(z))^2.
\end{aligned}
\end{equation}

The approximation in \eqref{8} stems from Taylor's approximation $\sqrt{1+x}\approx 1+x/2-x^2/8$, and the approximation in \eqref{10} can be applied when antenna plane dimensions are much smaller than the center distance between the transceiver and the receiver. The channel matrix in \eqref{4} with the approximated antenna distance is equivalent as 
\begin{equation}
\label{11}
\mathbf{H}\approx \frac{1}{4\pi D}\mathbf{F}_{\mathrm{RX}}{\mathbf{PF}_{\mathrm{TX}}}^*,
\end{equation}
where $\mathbf{F}_{\mathrm{RX}}\in \mathbb{C} ^{N\times N}$ and $\mathbf{F}_{\mathrm{TX}}\in \mathbb{C} ^{M\times M}$ are diagonal matrices representing the phase shifts caused by the receiver and transmitter independently. Specifically, 
\begin{equation}
\label{12}
\footnotesize
\begin{aligned}
&\mathbf{F}_{\mathrm{TX}}( m,m ) =\exp\{ -\frac{jk}{2D}[ -2x_0\varDelta _{xm}^{t}+( \varDelta _{xm}^{t} ) ^2-2y_0\varDelta _{ym}^{t}\\
&+( \varDelta _{ym}^{t} ) ^2-2z_0\varDelta _{zm}^{t}+( \varDelta _{zm}^{t}) ^2-\frac{ (\varDelta _{xm}^{t} x_0+ \varDelta _{ym}^{t}  y_0+\varDelta _{zm}^{t}z_0)^2}{D^2}]\}, 
\end{aligned}
\end{equation}
\begin{equation}
\label{13}
\footnotesize
\begin{aligned}
&\mathbf{F}_{\mathrm{RX}}( n,n ) =\exp \{ -\frac{jk}{2D}[ 2D^2+2x_0\varDelta _{xn}^{r}+( \varDelta _{xn}^{r} ) ^2+2y_0\varDelta _{yn}^{r}\\
&+( \varDelta _{yn}^{r} ) ^2+2z_0\varDelta _{zn}^{r}+( \varDelta _{zn}^{r} ) ^2-\frac{( \varDelta _{xn}^{r}x_0+ \varDelta _{yn}^{r}y_0+\varDelta _{zn}^{r}z_0)^2}{D^2}]\}, 
\end{aligned}
\end{equation}
and $\mathbf{P}$ is the channel matrix contributed by spatial multiplexing and is a non-diagonal matrix, given by 
\begin{equation}
\label{14}
\footnotesize
\begin{aligned}
&\mathbf{P}( n,m ) =\exp \{ \frac{jk}{D}[ \varDelta _{xn}^{r}\varDelta _{xm}^{t}+\varDelta _{yn}^{r}\varDelta _{ym}^{t}+\varDelta _{zn}^{r}\varDelta _{zm}^{t}\\
&-\frac{(\varDelta _{xm}^{t}x_0+\varDelta _{ym}^{t}y_0+\varDelta _{zm}^{t}z_0)( \varDelta _{xn}^{r}x_0+\varDelta _{yn}^{r}y_0+\varDelta _{zn}^{r}z_0 )}{D^2}]\}. 
\end{aligned}
\vspace{-0.2cm}
\end{equation}

Therefore, $\mathbf{G}$ can be written as follows
\begin{equation}
\small
\label{15}
\vspace{-0.2cm}
\begin{aligned}
    \mathbf{G}&=\frac{1}{\left( 4\pi D \right) ^2}\left[ \mathbf{F}_{\mathbf{RX}}{\mathbf{PF}_{\mathbf{TX}}}^* \right] ^*\mathbf{F}_{\mathbf{RX}}{\mathbf{PF}_{\mathbf{TX}}}^*\\
    &=\frac{1}{\left( 4\pi D \right) ^2}\mathbf{F}_{\mathbf{TX}}\mathbf{P}^*{\mathbf{PF}_{\mathbf{TX}}}^*.
\end{aligned}
\end{equation}

Because $\mathbf{F}_{\mathbf{TX}}$ is a diagonal matrix, it doesn't have any impact on the diagonalization of $\mathbf{G}${}. Let $\mathbf{R}=\mathbf{P}^*\mathbf{P}$ denote the channel gain matrix, which the rank of $\mathbf{R}$ being equal to that of $\mathbf{G}$\cite{MA2}. Therefore, by denoting 
$p(x)=\varDelta _{xv}^{t}-\varDelta _{xu}^{t} $, the $(u,v)$-th element of the channel gain matrices $\mathbf{R}$ is given by 
\vspace{-0.3cm}
\begin{equation}
\vspace{-0.2cm}
\label{16}
\footnotesize
\begin{aligned}
\mathbf{R}\left( u,v \right) &=\sum_{n=1}^N{\left[ \mathbf{P}\left( n,u \right) \right] ^*\mathbf{P}\left( n,v \right)}\,\,  \\
&=\sum_{n=1}^N\exp ( \frac{jk}{D}[ \varDelta _{xn}^{r}p(x) +\varDelta _{yn}^{r}p(y) +\varDelta _{zn}^{r}p(z) \\
&-\frac{( \varDelta _{xn}^{r}x_0+\varDelta _{yn}^{r}y_0+\varDelta _{zn}^{r}z_0 ) (p(x)x_0+p(y)y_0+p(z)z_0)}{D^2}.
\end{aligned}
\end{equation}

According to \eqref{TX} and \eqref{RX}, we denote the position indices of the $u$-th antenna, $v$-th antenna and $n$-th antenna as $(u_1,u_2)$, $(v_1,v_2)$ and $(n_1,n_2)$, respectively. Then substituting the antenna-specific coordinates, we expand and simplify \eqref{16}, shown as
\vspace{-0.2cm}
\begin{equation}
\label{17}
\vspace{-0.2cm}
\footnotesize
\begin{aligned}
&\mathbf{R}( u,v )=\sum_{n_1=1}^{N_H}\sum_{n_2=1}^{N_V}\exp  \{jk \{[\eta _{11} ( v_1-u_1 )+\eta _{12}( v_2-u_2 )]\\
&( n_1-\frac{N_H-1}{2} ) +[\eta _{21}( v_1-u_1 ) +\eta _{22} ( v_2-u_2 )]( n_2-\frac{N_V-1}{2} ) \} \},
\end{aligned}
\end{equation}
where 
\vspace{-0.2cm}
\begin{equation}
\label{18}
\footnotesize
\begin{aligned}
\eta _{11}&=d_{rh}d_{th}\frac{1}{D^3}(\sigma _1\left( \alpha _1,\alpha _2 \right) +{y_0}^2\sin \alpha _1\sin \alpha _2+x_0y_0\\
&\times\sin\mathrm{(}\alpha _1-\alpha _2)),
\end{aligned}
\end{equation}
\begin{equation}
\label{19}
\footnotesize
\begin{aligned}
\eta _{12}&=d_{rh}d_{tv}\frac{1}{D^3}(\sin \beta _1\sigma _2\left( \beta _1,\alpha _1,\alpha _2 \right) -z_0\cos \beta _1(x_0\cos \alpha _2\\
&+y_0\sin \alpha _2)),
\end{aligned}
\end{equation}
\begin{equation}
\label{20}
\footnotesize
\begin{aligned}
\eta _{21}&=d_{rv}d_{th}\frac{1}{D^3}(\sin \beta _1\sigma _2\left( \beta _2,\alpha _2,\alpha _1 \right) -z_0\cos \beta _1(x_0\cos \alpha _2\\
&+y_0\sin \alpha _2)),
\end{aligned}
\end{equation}
\begin{equation}
\label{21}
\footnotesize
\begin{aligned}
\eta _{22}&=d_{rv}d_{tv}\frac{1}{D^3}(\sin \beta _1\sin \beta _2\left( \sigma _1\left( \alpha _1+\frac{\pi}{2},\alpha _2+\frac{\pi}{2} \right) +\sigma _4(\alpha _1,\alpha _2 ) \right) \\
&+x_0z_0\sigma _3\left( \alpha _1,\alpha _2 \right) -y_0z_0\sigma _3\left( \alpha _1+\frac{\pi}{2},\alpha _2+\frac{\pi}{2} \right),
\end{aligned}
\end{equation}
with $\sigma _1( \alpha _1,\alpha _2 ) =D^2\cos\mathrm{(}\alpha _1-\alpha _2)-{x_0}^2\cos \alpha _1\cos \alpha _2$, $\sigma _2( \beta _1,\alpha _1,\alpha _2 ) =-D^2\sin\mathrm{(}\alpha _1-\alpha _2)+{x_0}^2\sin \alpha _1\cos \alpha _2-{y_0}^2\cos \alpha _1\sin \alpha _2-x_0y_0\cos\mathrm{(}\alpha _1+\alpha _2)$, $\sigma _3( \alpha _1,\alpha _2) =\sin \alpha _1\sin \beta _1\cos \beta _2+\sin \alpha _2\cos \beta _1\sin \beta _2
$, and $\sigma _4( \alpha _1,\alpha _2)=-{y_0}^2\cos \alpha _1\cos \alpha _2+D^2\cos \beta _1\cos \beta _2$$-{z_0}^2\cos \beta _1\cos \beta _2+x_0y_0\sin ( \alpha _1+\alpha _2 ) $.

The expression in \eqref{17}  can be simplified by using the geometric sum formula $\sum\nolimits_{n=0}^{M-1}{x^n=\left( 1-x^M \right) /\left( 1-x \right)}$  and the trigonometric identity $\sin \left( x \right) =\left( e^{jx}-e^{-jx} \right) /\left( 2j \right) 
$, which results in the following expression:
\begin{equation}
\label{22}
\footnotesize
\begin{aligned}
&\sum_{n=1}^N{[ \mathbf{P}( n,u ) ] ^*\mathbf{P}( n,v )}=\frac{\sin ( k( \eta _{11}( u_1-v_1 ) +\eta _{12}( u_2-v_2 )) \frac{N_H}{2})}{\sin ( k( \eta _{11}( u_1-v_1 ) +\eta _{12}( u_2-v_2 ) \frac{1}{2} ) )}\\
&\times \frac{\sin ( k( \eta _{21}( u_1-v_1 ) +\eta _{22}( u_2-v_2 ) ) \frac{N_V}{2} )}{\sin ( k( \eta _{21}( u_1-v_1 ) +\eta _{22}( u_2-v_2 ) ) \frac{1}{2})}.
\end{aligned}
\end{equation}

    To satisfy the channel orthogonality condition of \eqref{6}, the matrix elements represented by $\mathbf{R}$ as shown in \eqref{22} have to be equal to zero at non-diagonal positions. Then, the condition can be given by 
    \vspace{-0.2cm}
    \begin{equation}
\label{condition}
\small 
\mathbf{R}\left( u,v \right) =\sum_{n=1}^N{\left[ \mathbf{P}\left( n,u \right) \right] ^*\mathbf{P}\left( n,v \right)}=0\,\, \forall u\ne v=1,2,...,M.
\end{equation}
According to \cite{HG}, the orthogonality condition presented in \eqref{condition} admits a solution provided that at least one $\eta_{ab}(a,b={1,2})$ is equal to zero. 
\begin{corollary}

    When the transceiver planes are parallel and the antenna spacing is uniform with $\alpha_1,\beta_1,\alpha_2,\beta_2=0$ and $d_{th}=d_{tv}=d_{rh}=d_{rv}=d$, the antenna spacing can be expressed as 
\begin{equation}
\label{spacing}
\small
    d=\sqrt{\left| \frac{\lambda D^3}{N_H\left( x_0z_0 \right)} \right|}.
\end{equation}
\end{corollary}
\begin{IEEEproof}
    Based on the parameter conditions shown in Corollary 1, $\eta_{ab}$ can be simplified as follows:
\begin{equation}
    \footnotesize
    \eta _{11}=\frac{1}{D^3}\left( D^2d^2-{x_0}^2d^2 \right),
\end{equation}
\begin{equation}
\vspace{-0.2cm}
    \footnotesize
    \eta _{12}==-x_0z_0d^2\frac{1}{D^3},
\end{equation}
\begin{equation}
\vspace{-0.2cm}
    \footnotesize
    \eta _{21}=-x_0z_0d^2\frac{1}{D^3},
\end{equation}
\begin{equation}
    \footnotesize
    \eta _{22}=\frac{1}{D^3}\left( D^2d^2-{z_0}^2d^2 \right).
\end{equation}
Based on the system model shown in Fig. 1, $\eta _{11}\approx0$. Besides, the function $\sin ( \pi N_H\eta _{12}( u_2-v_2 ) /\lambda ) /\sin ( \pi \eta _{12}( u_2-v_2 ) /\lambda) 
$ with $\lambda=2\pi/k$ is a periodic function with period $\lambda/\eta _{12}$ and it has a zero in $q/\eta _{12}$ for $q \in(\lambda/N_H,(N_H-1)\lambda/N_H)$. Then, we can ensure the channel orthogonality condition is fulfilled when $|N_H\eta _{12}/\lambda|=1$. Then, the expression of antenna spacing can be proved. 
\end{IEEEproof}
\begin{remark}
    In practical research, solving the solution of the channel orthogonality condition is too complicated, so we only perform a simple analysis here for the state where the rotation angles $(\alpha_1,\beta_1,\alpha_2,\beta_2)$ are all zero. In the case of considering arbitrary rotation angles, in order to simplify the calculation, we use the rank of the $\mathbf{R}$ as the optimization index. 
\end{remark}

\vspace{-0.7cm}
\subsection{Differential Evolution Algorithm for ROMA}
Based on the channel correlation analysis, we propose a framework, shown in Algorithm 1, for the optimization of movable antenna configurations to calculate the optimal rotation angles of the transmitting and receiving surface panels for the positional variations brought about by the train running in an HSR scenario. We use the rank of the channel correlation matrix as an optimization variable. It reflects the number of independently transmitted signals that the channel can support, which can directly affect the channel capacity and is an important indicator of channel performance. The optimization problem can be written as
\vspace{-0.1cm}
\begin{equation}
\label{28}
\small
\begin{aligned}
     \left( \hat{\alpha}_1,\hat{\beta}_1,\hat{\alpha}_2,\hat{\beta}_2 \right) &=\underset{\alpha _1,\beta _1,\alpha _2,\beta _2}{\mathrm{arg}\max}\left\{ \mathrm{rank}\left( \mathbf{R} \right) \right\} \\
     \mathrm{subject\,to}\,\,\,&{\alpha}_1,{\alpha}_2\in[-\pi/2,\pi/2], \\
     &{\beta}_1,{\beta}_2\in[0,\pi/2], 
\end{aligned}
\end{equation}
where the range of values for the rotation angle is based on the physical scenario in the system model.

\begin{algorithm}[!h]
    \caption{Differential Evolution Algorithm for ROMA}
    \label{alg: AOA}
    \renewcommand{\algorithmicrequire}{\textbf{Input:}}
    \renewcommand{\algorithmicensure}{\textbf{Output:}}
    \begin{algorithmic}[1]
        \REQUIRE The range of values for $\alpha_1,\beta_1,\alpha_2,\beta_2$, population number $num$, variable dimension $de$ in each individual, maximum number of evolutionary generations $gen$, initial variation operator $F_0$, and crossover operator $C_R$.

        \STATE $\mathbf{Initialisation:}$ Assign each dimension to each individual in the population randomly within a range of values, shown as $\mathbf{x} \in \mathbb{R} ^{de\times num} $.

        \FOR {$i$=1,2,..., $num$}
        \STATE Evaluation of the $i$-th individual in the original population according to \eqref{28}.
        \ENDFOR
        \STATE Tracking optimal individuals in a population $trace(1)$.
        \FOR{$g$=1,2,..., $gen$}
\STATE Dynamically set the variation operator according to the number of iterations: $lamb=exp(1-gen/(gen+1-g))$, $F=F_02^{lamb}$.
\FOR{$i$=1,2,..., $num$}
\STATE Randomly take three unequal integers $r_1, r_2, r_3\ne i$ within the population size. Then, calculate the variant individual as:
$\mathbf{y}(:,i)=\mathbf{x}(:,r_1)+F(\mathbf{x}(:,r_2)-\mathbf{x}(:,r_3))$, and verify the range of values.
\ENDFOR
\STATE Iterate through each individual of population $\mathbf{x}$, generating a random probability each time, and if the random probability is greater than the crossover probability, exchange the individuals of population $\mathbf{x}$ with the individuals of population $\mathbf{y}$ at the corresponding index.

\FOR {$i$=1,2,..., $num$}
        \STATE Evaluation of the $i$-th individual in the population $\mathbf{x}$ according to \eqref{28}.
        \ENDFOR
        \STATE Tracking optimal individuals in $trace(g+1)$.
        \ENDFOR              \ENSURE The optimal individual $trace(gen+1)$; the optimal rotation angle in this individual $\hat{\alpha}_1,\hat{\beta}_1,\hat{\alpha}_2$, and $\hat{\beta}_2$.
    \end{algorithmic}
    \vspace{-0.1cm}
\end{algorithm}

In Algorithm 1, we use adaptive differential evolution\cite{ref3} to solve the above optimization problem. The genetic difference method is an efficient global optimization algorithm, which is widely used because it is less affected by parameters and is suitable for solving the optimal value in multi-dimensions. The algorithm performs parallel computation from a population, where each individual in the population corresponds to a solution vector represented in the optimisation problem, generates new individuals through mutation, hybridization and selection operations, and evaluates and compares the different individuals using a greedy criterion, which guides the search process towards the optimal solution. At the same time, we use a dynamically changing mutation factor $F$ to improve the quality of individual searches.
\vspace{-0.3cm}
\section{Simulation Results and Discussions}
In this section, we provide the simulations to demonstrate the performance of the ROMA-HSR XL-MIMO framework. We consider a downlink XL-MIMO system where the transmitter and receiver are equipped with the same square UPA. We denote the carrier frequency $f_c=20\,\mathrm{GHz}$, the wavelength $\lambda_c=c/f_c$, the train speed $v=350\,\mathrm{km/h}$, and $\sum\nolimits{P_i=P_{max}}$, where $P_{max}$ is the maximum transmit power. Then, we consider each transmitted data stream with the same power and the signal-to-noise ratio (SNR) is $15\,\mathrm{dB}$. Besides, the relative height between the transmitter and receiver is $|z_0|=10\,\mathrm{m}$, and the horizontal distance from BS to train is $|y_0|=4\,\mathrm{m}$.

To quantity the accuracy of the localization model, we use the normalized mean square error (NMSE) as the metric:
\vspace{-0.2cm}
\begin{equation} 
NMSE=\frac{\sum\nolimits_{t=0}^T{\left| \left( \theta _t,R_t \right) -\left( \hat{\theta}_t,\hat{R}_t \right) \right|^2}}{\sum\nolimits_{t=0}^T{\left| \left( \theta _t,R_t \right) \right|^2}}.
\end{equation}

\begin{figure}[!t]
\vspace{-0.3cm}
\centering
\includegraphics[scale=0.4]{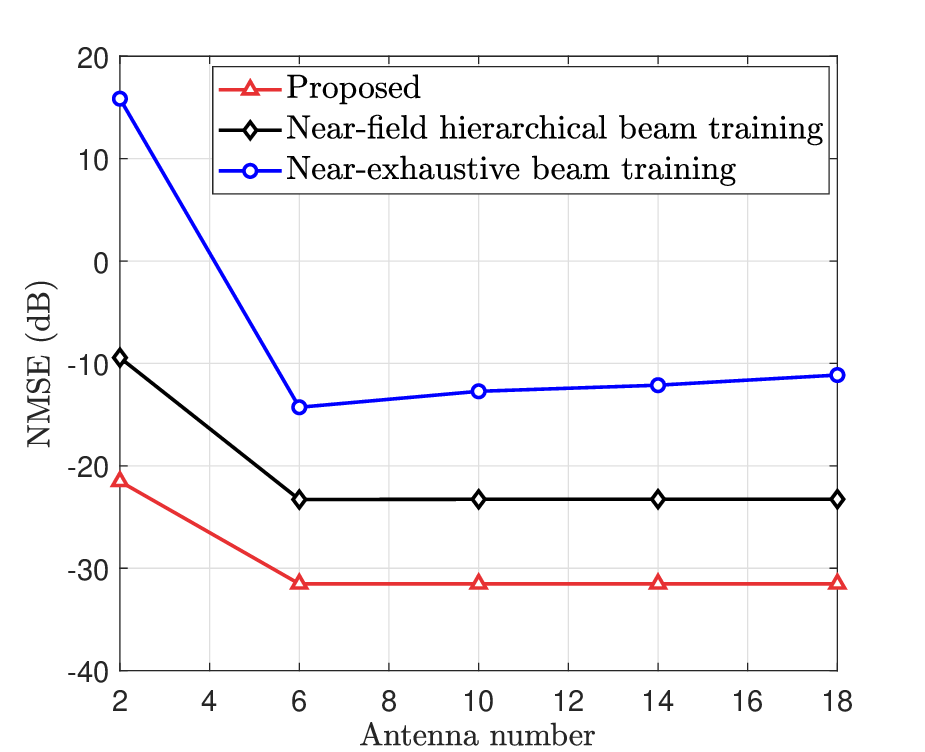}%
\label{second_second_case}
\vspace{-0.3cm}
\caption{Localization NMSE against the antenna number of the transmitter and receiver with different near-field beam train algorithms.}
\label{fig_sim}
\vspace{-0.5cm}
\end{figure}
In Fig. 2, we simulate the accuracy of the localization model by leveraging the characteristics of near-field beam training for train relay localization. The train moves along a trajectory within a $2000\,\mathrm{m}$ radius centered on the base station at a constant speed, with the base station recording the position of the train relay every second. By comparing our proposed localization model with two widely used near-field beam training algorithms, we demonstrate that our model significantly reduces the localization error, enabling more precise positioning of the train relay. Additionally, it is observed that increasing the number of antennas reduces the error in the near-field beam training algorithms. This improvement arises because a larger antenna array expands the near-field range, assuming fixed antenna spacing. 

\begin{figure}[!t]
\vspace{-0.3cm}
\centering
\includegraphics[scale=0.4]{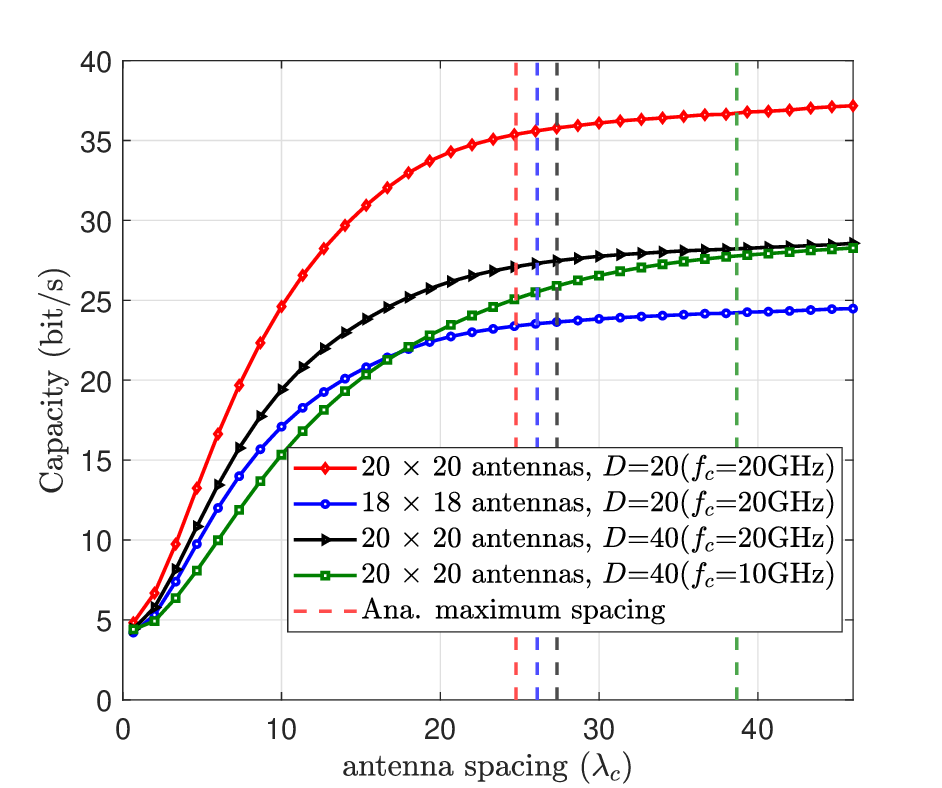}%
\label{second_second_case}
\vspace{-0.2cm}
\caption{Capacity versus normalized antenna spacing for different antenna numbers, physical distances and carrier frequencies.}
\label{fig_sim}
\vspace{-0.3cm}
\end{figure}
In Fig. 3, we compare the relationship between antenna spacing and channel capacity by varying antenna configurations, assuming the receivers and transmitters are parallel to each other. As shown by the solid line, for a given number of antennas, the channel capacity increases with antenna spacing until it converges. The larger the number of antennas, the higher the maximum channel capacity that can be achieved. Additionally, the greater the distance between the receiver and transmitter, the lower the maximum achievable channel capacity. When other parameters remain constant, a decrease in carrier frequency results in a slower convergence rate, though it does not affect the convergence value. The dashed line represents the optimal antenna spacing, calculated using the analytic expression in \eqref{spacing}, which closely approaches the minimum antenna spacing required for channel capacity convergence. This confirms the accuracy of Corollary 1.

\begin{figure}[!t]
\centering
\includegraphics[scale=0.4]{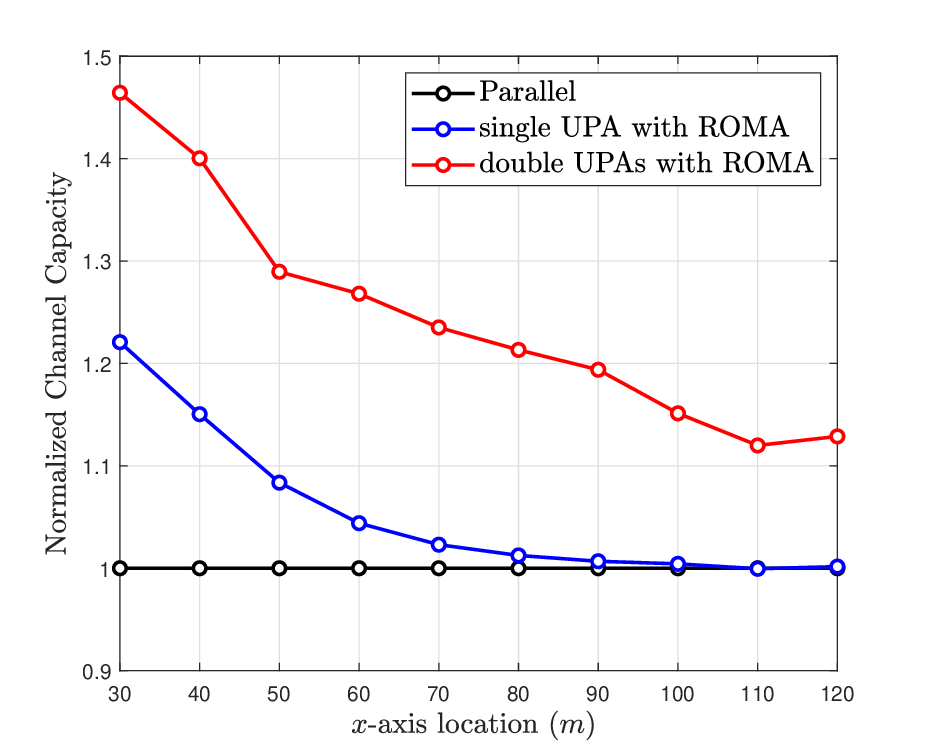}%
\label{second_second_case}
\vspace{-0.3cm}
\caption{Normalized channel capacity versus the $x$-axis location for three cases: parallel receiver and transmitter with FPA, one-sided plane with ROMA, and both planes with ROMA.}
\label{fig_sim}
\vspace{-0.5cm}
\end{figure}
In Fig. 4, we compare the relationship between channel capacity and location under three different conditions, with $20 \times 20$ movable antennas configured at both the transmitter and receiver. For the analysis, we set $F_0 = 0.5$, $C_R = 0.2$, and assume the train travels at a constant speed to various positions. The x-axis positions correspond to the train’s relay positions, as $y_r$ and $z_r$ remain fixed. The channel capacity is normalized by using the scenario where both the receivers and transmitters are parallel and fixed as the reference. The results indicate that system performance improves when both the receivers and transmitters can rotate within a certain range. Specifically, configurations where the panels can rotate outperform the fixed setup. For instance, when the train reaches the x-axis position of $40\,\mathrm{m}$, the channel capacity of both bilateral XL-MIMOs equipped with ROMA is $1.4$ times greater than that with FPA, and the single-sided XL-MIMO with ROMA yields a channel capacity $1.15$ times higher than that with FPA. However, this performance gain diminishes as the distance increases, due to the growing influence of large-scale fading.
\vspace{-0.3cm}
\section{Conclusion}
In this correspondence, we investigate the performance of the XL-MIMO system with ROMA in the HSR communication scenario, focusing on channel correlation and channel capacity. We first adopt a mobility-aware near-field beam training approach to locate the train. Subsequently, by analyzing the channel correlation matrix, we derive the channel orthogonality condition and determine the optimal antenna spacing expression under specific conditions. Furthermore, we employ the differential evolution method to optimize the rotation angles of the receiver and transmitter panels. Finally, simulation experiments are conducted to validate the performance gains achieved by ROMA in the HSR communication system. In future work, we will continue to explore the performance and key technologies of the XL-MIMO system with ROMA in more complex HSR scenarios.

%

\newpage
 
\vspace{11pt}

\vspace{11pt}

\vfill

\end{document}